\documentclass{article}

\usepackage{PRIMEarxiv}

\usepackage[utf8]{inputenc} % allow utf-8 input
\usepackage[T1]{fontenc}    % use 8-bit T1 fonts
\usepackage{hyperref}       % hyperlinks
\usepackage{url}            % simple URL typesetting
\usepackage{booktabs}       % professional-quality tables
\usepackage{amsfonts}       % blackboard math symbols
\usepackage{nicefrac}       % compact symbols for 1/2, etc.
\usepackage{microtype}      % microtypography
\usepackage{lipsum}
\usepackage{fancyhdr}       % header
\usepackage{graphicx}       % graphics
\graphicspath{{media/}}     % organize your images and other figures under media/ folder

\title{Gerald Edelman's Steps Toward a Conscious Artifact}

\author{
  Jeffrey L. Krichmar \\
  Department of Cognitive Sciences \\
  Department of Computer Science \\
  University of California, Irvine \\
  Irvine, CA 92697-5100\\
  \texttt{jkrichma@uci.edu} \\

}

\begin{document}
\maketitle

\begin{abstract}
In 2006, during a meeting of a working group of scientists in La Jolla, California at The Neurosciences Institute (NSI), Gerald Edelman described a roadmap towards the creation of a Conscious Artifact.  As far as I know, this roadmap was not published. However, it did shape my thinking and that of many others in the years since that meeting. This short paper, which is based on my notes taken during the meeting, describes the key steps in this roadmap. I believe it is as groundbreaking today as it was more than 15 years ago.
\end{abstract}

% keywords can be removed

\section{Introduction}

In February of 2020, I participated in the "On Consciousness" podcast with Bernie Baars and David Edelman. We talked about my work at The Neurosciences Institute (NSI) in La Jolla, California on the Darwin series of Brain-Based Devices, as well as my current research in neurorobotics. Unsurprisingly, the conversation turned to consciousness.  I happened to mention that a page from my old lab notebook, which is pinned to a bulletin board in my office at UC Irvine, outlines a roadmap towards the creation of a \textit{Conscious Artifact}. The key steps in this roadmap were laid out by Gerald Edelman, who was the director of the NSI at the time I was a research fellow there.

\begin{figure}[ht!]
\begin{center}
\includegraphics[width=0.95\linewidth]{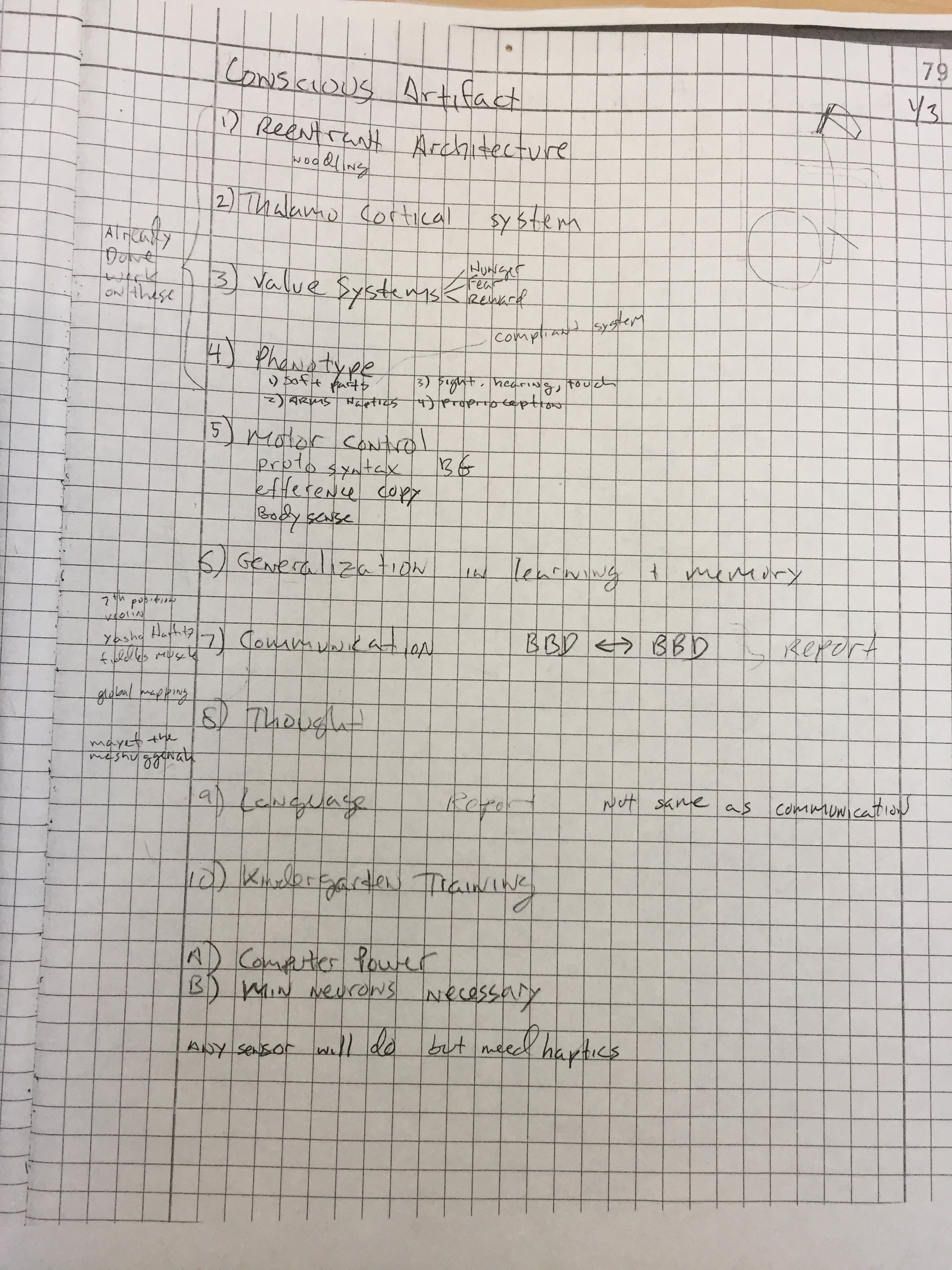}
\end{center}
\caption{Copy of the roadmap for a Conscious Artifact from my lab notebook in 2006.}
\label{fig:ca}
\end{figure}

Several months later, alert listener Grant Castillou contacted me about the podcast. In his email he wrote, "I have read many of Dr. Edelman's works, but never come across anything like this before". At that time, my university was in lockdown due to the COVID-19 pandemic and I did not have access to my office. Recently, though, I was in my office and found the roadmap (see Figure \ref{fig:ca}). This brief paper describes the roadmap and my recollections of the event and the thinking behind it more than 15 years ago.

\section{Background and History}
\label{sec:bh}
It is necessary to provide some background on what led to this event.  Of course, it was a long time ago, so I may not have all the facts and dates exactly right. It is well known that memory is labile; time and the accumulation of experience can lead to false memories. Keeping that caveat in mind, what follows is my best recollection of the event and what was proposed. 

Although the date in the upper right corner of my notebook looks like January 3rd, I think this is either a misprint or my bad handwriting. I believe the event was shortly after the Society for the Study of Artificial Intelligence and Simulation of Behaviour (AISB) meeting held in Bristol UK in April of 2006. I participated in the GC5: Architecture of Brain and Mind symposium and also attended the symposium on Integrative Approaches to Machine Consciousness.

When I returned to the NSI, I was asked to give a briefing on the meeting at the Institute's Fellows' Symposium. I described what I saw and, in particular, the current state of machine consciousness and artificial intelligence. 

Later that day, we had our weekly "Build A Brain" meeting. The "Build A Brain" group comprised an eclectic mix of theoreticians, engineers, and neuroscientists. Every week we would discuss the progress on our brain-based devices and how we could test theories of neuroscience using those simulations. Although Gerald Edelman had given much thought to the topic of consciousness, up until that point he had not wanted to construct a conscious machine. In fact, he either did not think it was possible or thought it was not a worthwhile endeavor.

However, at this particular meeting, I think my report on AISB'06 compelled him to state, "I think the time has come to build a conscious artifact." My notes from that meeting are shown in Figure \ref{fig:ca}. After the meeting, Eugene Izhikevich, who was also present, told me to save this page because it might be important. So, I made an extra copy of the page for myself, and it eventually followed me to UC Irvine, where it now occupies a corner of my bulletin board.

The rest of this paper describes, to the best of my recollection, my interpretation of this roadmap toward a \textit{Conscious Artifact}. 

\section{Roadmap to a Conscious Artifact}
\label{sec:roadmap}

Table \ref{tab:ca} is re-typed from my notes in Figure \ref{fig:ca}. Anyone who had the good fortune to be in a meeting with Gerald Edelman would know that thoughts peppered listeners in rapid fire. Or, as he would say, a "drink from a fire hose."  Without notes or other written prompts, Edelman rattled off 10 key steps toward a Conscious Artifact. I tried as best as I could to keep up. The steps were presented roughly in order of how one would proceed to construct a Conscious Artifact. The first four steps were areas in which the team had made some progress. The remaining steps were areas that needed further investigation. I would say that, to date, none of the steps listed have been adequately addressed. However, the purpose of this paper is not to review the current state of the art, but rather to describe these steps as best I can. 

\begin{table}
 \caption{Conscious Artifact}
  \centering
  \begin{tabular}{ll}
    \toprule
    Step     & Description      \\
    \midrule
    1)  & Reentrant Architecture      \\
    2)  & Thalamo-Cortical System    \\
    3)  & Value Systems        \\
    4)   & Phenotype \\
    5)   & Motor Control \\
    6)   & Generalization in Learning and Memory \\
    7)   & Communication \\
    8)   & Thought \\
    9)   & Language \\
    10)  & Kindergarten \\
    A)  & Computer Power \\
    B)  & Minimum Neurons Necessary \\
    Note:    & Any Sensor will Do But Need Haptics \\
    \bottomrule
  \end{tabular}
  \label{tab:ca}
\end{table}

\subsection{Reentrant Architecture}
A major component of Edelman's Neural Darwinism and Theory of Neuronal Group Selection (TNGS) was reentrant signaling \cite{edelman1987}. Reentrant, because it was different than feedback signals. Neuronal groups were bi-directionally linked with synaptic connections. Different groups with specific features could share information with these connections. Experience-dependent plasticity selected associations among these groups. These global mappings (see margin of Figure \ref{fig:ca}) ultimately gave rise to perceptual categories and action plans.

The theoretical neuroscientists working at the NSI had created a number of models to test these ideas, both in simulation and instantiated on physical robotic platforms \cite{seth2004, wray1996}. With these models, they were able to show feature mapping, binding through synchrony, memory recall, and other brain properties. These were colorfully termed, "noodling," as can be read in my notes.

\subsection{Thalamo-Cortical System}

In a series of books, Edelman described his theory of consciousness, which was based on the TNGS \cite{edelman1990,edelman1993}. A key to this theory, as enumerated in some of his later publications, was the concept of a Dynamic Core \cite{edelman2000}. The Dynamic Core was essentially reentrant signalling between the thalamus and the neocortex.  The dynamics of the Dynamic Core were necessary to produce conscious thought or higher order consciousness.

At the time of the meeting, researchers at the NSI were developing extremely detailed computational models of the thalamus and  neocortex.  These models showed sleep wake cycles, as well as other other brain rhythms observed during conscious thought \cite{lumer1997a,lumer1997b}. In one such model, Eugene Izhikevich and Gerald Edelman were able to show the formation of neuronal groups due to plasticity and dynamic neuronal activity \cite{izzy2008}.

\subsection{Value Systems}

Value systems are neural structures that are necessary for an organism to modify its behavior based on the salience or value of an environmental cue. The value system in a brain-based device is analogous to neuromodulatory systems in that its units show phasic responses when activated and its output acts diffusely across multiple pathways to promote synaptic change.

Value systems were first explored by Karl Friston in theoretical work conducted at the NSI when the Institute was located on the campus of The Rockefeller University in New York City \cite{friston1994}. Every Brain-Based Device was equipped with a value system for shaping behavior and building associations between predicted value and observed value \cite{krichAlife2005}. Edelman noted that value could signal hunger, fear, and reward, among other signals salient to the behaving agent.

\subsection{Phenotype}

In the case of the robotic platforms created at the NSI, the phenotype consisted of the physical design of a given device. In particular, it was the shape or morphology of the robot, as well as the layout and type of on-board sensors. The Darwin series of automata had a basic shape;  notably, all were wheeled robots. But despite their relatively simple external design, they had a wide range of sensors for hearing, vision, taste, and touch. Later versions of the Darwin automata included omni-wheels, as well as articulated arms and legs.

Interestingly, Edelman noted that the phenotype needed to be compliant and must necessarily include proprioception. Compliance through soft materials and elasticity is an important property of biological systems. Proprioception would, Edelman believed, lead to a notion of self and body awareness.  These phenotypes were not a part of our design at the time and needed to be further investigated through incorporation into additional robotic platforms. Also notable is the fact that Edelman came back to the importance of touch toward the end of the meeting. See the last line of Figure \ref{fig:ca} and Table \ref{tab:ca}, where I quoted him saying "any sensor will do, but need haptics."

\subsection{Motor Control}

Edelman's theory of consciousness was very much tied to behavior and intended actions. Therefore, in his mind (and presumably those who attended the meeting) motor control was an important step towards creation of a Conscious Artifact. In particular, Edelman mentioned efference copy and body sense. After every action, a copy of the motor command is fed back to the nervous system. This is called a "motor efference copy". The brain uses the motor efference copy to check if the action  generated yields the expected sensory stimuli and expected body position. In this way, the agent might produce a body sense.

It is also notable that Edelman singled out the Basal Ganglia (BG in my notes) as an important aspect of motor control and key functional anatomical feature of consciousness.  Therefore a key step towards achieving a conscious artifact is incorporating Basal Ganglia function. The Basal Ganglia selects actions and generates putative motor sequences. In my notes, I wrote "proto syntax".  Edelman thought that action selection and the sequence of actions was a first step towards language. He was a strong proponent of the idea that language was rooted in action and motor control.

\subsection{Generalization in Learning and Memory}

Around this time, researchers at the NSI were constructing sophisticated models of hippocampus and long-term memory \cite{fleischer2007,krichPNAS2005}. However, these models were brittle.  They suffered from an inability to transfer information from one task to another, as well an incapacity for generalization. I find it interesting that, to this day, transfer learning and generalization continue to limit AI systems. Moreover, the ability to learn over a lifetime of experience remains beyond the reach of current artificial systems. 

\subsection{Communication}

Critically important to demonstrating a Conscious Artifact would be some form of accurate report. Before language, a report could be realized by communication between Brain-Based Devices (BBD <--> BBD, in Figure \ref{fig:ca}). By reporting its intentions and state to another agent, the agent is showing a degree of self-awareness. It is also interesting to speculate that consciousness, in particular self-awareness (or higher-order consciousness), might not be observable in a single agent. Rather, it might require social interaction.
 
 \subsection{Thought}
 
 Unfortunately, I don't have much to add to this step. I can only guess that here, Edelman was alluding to mental simulation and imagination. This is a prediction of the dynamic core theory, which was briefly described in the Thalamo-Cortical section and in more detail in Edelman's writings.
 
 \subsection{Language}
 
 Again, report is brought up as an important step toward creating a Conscious Artifact.  However, Edelman makes it clear that language is distinctively more sophisticated than communication, as described in the Communication step.  Language is nuanced, suffused as it is with emotion, thought, intention, and action.  It is safe to say that Edelman was thinking that language, as instantiated in a Conscious Artifact, would be something far beyond Natural Language Processing or simply passing a Turing test. An accurate report via language would need to demonstrate that the agent had a form of higher order consciousness closely tied to body sense and a self.
 
 \subsection{Kindergarten}
 
 Similar to Turing's theory and the field of developmental robotics, Edelman proposed that to achieve all of the above, the Conscious Artifact would need to be subjected to a curriculum of sorts. It was too much to load these characteristics upon initialization of a given simulation.  And more importantly, consciousness is tied to the individual's experience. This is where learning and memory from experience becomes critically important. Furthermore, communication and language are necessary in order to interact with a teacher or caretaker, not to mention, report one's intentions and state.
 
 \subsection{Other Notes}
 
 Towards the end of the meeting, some pragmatic issues were brought up. One limiting factor at the time was computational power, which remains a constraint today.  I am unsure what is meant by the "minimum neurons necessary", which I quoted in my notebook. It could be taken to be tantamount to computing power. Or it could be related to sparseness and energy. The metabolic constraints imposed by biology was a topic that Edelman often revisited during internal meetings at the NSI.

\section{Hidden Gems}

Anyone who had the experience of interacting with Gerald Edelman will tell you that conversations were generously sprinkled with anecdotes and a seemingly infinite supply of jokes. He was a consummate storyteller. Given any context, he would suddenly and unexpectedly tell a joke that not only made a point with great clarity, but was also delivered like a seasoned standup comedian. This meeting was no different in content or character.

During such meetings, I would make notes in the margins to remind me of the particular story or joke.

The first note is "7th position violin" and "Yasha Haifitz fiddles muscle". Edelman was an accomplished violinist. It was said that he could have had a career as a concert violinist. The 7th position is the most advanced position of the hand; necessary for generating the highest notes on a violin. According to members of Edelman's family, his mother wanted him to be a doctor. As the story goes, when he expressed the desire to become a concert violinist, his mother said, "I have two words for you: Jascha Heifetz". Heifetz was one of the greatest violinists of all time. This quickly ended the conversation. Edelman realized he could never be on a par with Heifetz and decided to go into medicine instead. This, of course, led to an accomplished research career and a Nobel Prize in Physiology or Medicine for deciphering the protein structure of the antibody (i.e., gamma gobulin or IgG). I believe the discussion of seventh position during the meeting had to do with motor control sequences and converting those sequences into muscle memory. In any case, it nicely follows from: 5) Motor Control to 6) Generalization in Learning and Memory.

I can only guess what "global mapping" meant. It certainly fits with the idea of reentrant architecture and the dynamic core.

Lastly, "Mayef the Meshuggenah" is an element of one of Edelman's classic jokes. Though the phrase certainly alludes to Albert Mayev, Edelman's violin teacher, who was good friends with a number of the great violinists of the mid-20th Century, including Heifetz and Nathan Milstein. Sadly, I cannot remember the details. It is not something that you can Google to understand the precise context. Perhaps one of my colleagues can recall the details of this joke or story.

\section{Final Thoughts}

I strongly believe that the steps laid out on that day in 2006, comprise the correct roadmap toward creation of machine consciousness. With this in mind, I believe it is important to share these steps with others. I have tried to follow them in my own research. But it is important to temper expectations here; each of these steps is a career in itself. For example, I decided to focus on "3) Value Systems" when I transitioned from the NSI to UC Irvine in 2008 \cite{krich2008}.  I am still researching this fascinating topic \cite{avery2017,krich2013}. Along the way, I have turned my attention to other steps on the list, but certainly have not addressed all the steps listed in Figure \ref{fig:ca} and Table \ref{tab:ca}.

After I left the NSI, Edelman, along with Yanqing Chen, Jason Fleischer, Joe Gally, Jeff McKinstry, and others, began working in earnest on a Conscious Artifact project. Among other things, they made notable progress on sequence learning and mental imagery \cite{mck2013,mck2016}.

Nevertheless, I think I can safely say that no one has put all these steps together within one system. I am hopeful that someone, either in my generation or the next, will one day be able to achieve this lofty goal.

\section*{Acknowledgments}
I would like to thank Bernie Baars and David Edelman for the discussion that brought this roadmap to mind, as well as for many interesting conversations. David was extremely helpful with editing and filling in details. A big shout out goes to Grant Castillou for bringing this to my attention and leading me to put this in writing so that it could be shared with others.  Lastly, I would like to thank the "Build A Brain" group at The Neurosciences Institute for generating these ideas, for the progress toward a Conscious Artifact, and for many fond memories.

%Bibliography
\bibliographystyle{unsrt}  
\bibliography{conscious}

\end{document}